\def\jobname{aa55255-25}

\documentclass{aa} 

\usepackage{graphicx}
\usepackage{txfonts}
\usepackage{lipsum}
\usepackage{subcaption}         
                                
\usepackage{lscape}             
                                
\usepackage{placeins}

\usepackage{natbib}
\bibpunct{(}{)}{;}{a}{}{,}
\begin{document}

   \title{Weak or strong: coupling of mixed oscillation modes on the red-giant branch}

   \author{T. van Lier\inst{1}\fnmsep\inst{2}
        \and J. Müller\inst{1}\fnmsep\inst{2}
        \and S. Hekker\inst{1}\fnmsep\inst{2}
        }

   \institute{Heidelberg Institute for Theoretical Studies, Schloss-Wolfsbrunnenweg 35, D-69118 Heidelberg, Germany
            \and Zentrum für Astronomie Heidelberg (ZAH/LSW), Heidelberg University, Königstuhl 12, D-69117 Heidelberg, Germany}

 
  \abstract
   {The high precision of recent asteroseismic observations of red-giant stars has revealed the presence of mixed dipole modes in their oscillation spectra. These modes allow for a look inside the stars. Among the parameters used to characterize mixed modes is the coupling strength $q$, which is sensitive to the stellar structure in the evanescent zone near the bottom of the convective envelope.}
   {The aim of this work is to probe the validity of the weak and strong coupling approximations, commonly used to calculate $q$, during stellar evolution along the red-giant branch (RGB).} 
   {To test the approximations empirically, we calculate $q$-values in both, the weak and strong limit for stellar models on the RGB and compare them to the coupling derived from the mixed mode frequency pattern obtained from numerical solutions to the oscillation equations.}
   {We find good agreement with the strong coupling approximation on the early RGB, when the evanescent zone lies in the radiative layer right above the hydrogen-burning shell; and with the weak coupling approximation once the evanescent zone is situated in the convective envelope. This is consistent with earlier studies. Additionally, we find that it is viable to use the weak coupling approximation as an estimate for $q$ in the intermediate regime, in the mass range considered in this work ($1.00\,\mathrm{M}_\odot\leq M\leq 2.00\,\mathrm{M}_\odot$).}
   {The width of the evanescent zone serves as a good measure for which approximation to use. The serendipitous alignment of the weak coupling approximation with the observable $q$ in the regime where neither approximation is expected to be valid simplifies the asymptotic calculation of mixed mode properties.}


   \maketitle

\section{Introduction}\label{s:introduction}

Conventional observations limit astronomers to studying the properties of the very outer layers of stars. Asteroseismology opens a window to look inside by studying periodic perturbations to the equilibrium stellar structure which can be observed as variations in the brightness of a star. With different oscillation modes being sensitive to different depths beneath the stellar surface, this allows for the investigation of features such as structure parameters, rotation rates or magnetic field strengths across the stellar profile (see \citet{lit:Hekker2017} for a review specialized on red-giant stars). To extract these features from observations, it is necessary to understand the interaction of the different modes with each other and how it affects observables such as the eigenfrequencies. The coupling of mixed modes in red-giant stars we study here is such an interaction which allows to study stellar parameters for the core and the envelope of these stars differentially. 

The oscillations of solar-like oscillators (including red-giants) are stochastically driven by the transfer of energy from turbulent convective motions in the envelope to resonant modes with frequencies similar to the inverse of the convective timescale. This leads to excitation within a Gaussian power envelope around a frequency of maximum oscillation power~$\nu_\mathrm{max}$, which is related to the star's mass~$M$, radius~$R$ and effective temperature~$T_\mathrm{eff}$ \citep[e.g.\,][]{lit:Kjeldsen1995}:
\begin{align}
    \nu_\mathrm{max}\propto\frac{M}{R^2\sqrt{T_\mathrm{eff}}}\,.
    \label{eq:numax_scaling}
\end{align}
Using this scaling relation and solar reference values, $\nu_\mathrm{max}$ can be used as a measure for the global stellar $M$ and $R$.

Red-giant stars have two regions in which perturbations to the equilibrium stellar structure can oscillate: a gravity(g)-mode cavity in the core, where buoyancy acts as the restoring force, and a pressure(p)-mode cavity in the envelope, where oscillations behave like sound waves. In between lies a layer in which perturbations decay exponentially, the so-called evanescent zone \citep[e.g.\,][]{lit:Hekker2017}. During evolution along the red-giant branch (RGB), the evanescent zone first lies in the radiative layer above the hydrogen-burning shell and then later moves further outwards into the bottom of the convective envelope \citep[e.g.\,][]{lit:vanRossem2024,lit:Pincon2020}.

Although different in nature, internal gravity-modes and pressure-modes can resonate at similar frequencies in red-giant stars. This allows for coupling of the two cavities, leading to mixed modes which have a dual character: they oscillate as g-modes in the core and as p-modes in the envelope of the star. Just like for the coupling of harmonic oscillators, this affects the eigenfrequencies of the modes compared to their values for separate cavities. The resulting mixed mode frequency pattern was first asymptotically described by \citet{lit:Shibahashi1979}, who also introduced the notion that the strength of the coupling is governed by the properties of the evanescent zone. This region between the hydrogen-burning shell and the bottom of the convective envelope is otherwise inaccessible, since it neither significantly contributes to the star's energy production, nor is it in chemical equilibrium with the surface by convective mixing.

In the limit of weak coupling, i.e.\,a relatively wide evanescent zone, \citet{lit:Shibahashi1979} derived a first expression for the coupling strength $q$ which confines its value to $q\leq1/4$. Observations of red-giants, however, revealed that this upper limit is exceeded in some of the stars, especially in the red clump \citep[e.g.\,][]{lit:Mosser2012}. In an effort to explain the observed high values of the coupling strength \citet{lit:Takata2016a} derived an alternative prescription in the contrasting limit of strong coupling, i.e.\,a relatively narrow evanescent zone. \citet{lit:Mosser2017} also observed sub-giant and early-RGB stars with $q>1/4$, which motivated that strong coupling is also relevant for these evolutionary stages.

Since then, the validity of the two approximations has been widely discussed in the literature, e.g.\,by comparing their predictions to ensemble observations \citep{lit:Pincon2020} as well as to a fit of the pattern of g-modes coupling to each p-mode of models \citep{lit:Jiang2022a}. These studies find that the strong coupling approximation is valid as long as the evanescent zone lies in the radiative layer directly above the hydrogen-burning shell, while the weak coupling approximation is applicable once the evanescent zone is located in the convective envelope; with need for a prescription for the transition in between. The evolution of the evanescent zone properties in models has been studied in detail by \citet{lit:vanRossem2024}.

By comparing the coupling derived from the pattern of numerically computed mixed modes of stellar models across multiple acoustic orders to the weak and strong coupling calculated for the same models, we test these findings empirically along stellar evolution on the RGB in a setting more closely related to an observational approach. \citet{lit:Jiang2020} showed that the coupling strength derived from the mixed mode pattern varies with frequency. When working with observations, however, an order-wise fitting as performed by \citet{lit:Jiang2022a} is not feasible, because only a limited fraction of the mixed modes is detectable and therefore the statistical information in one acoustic order does not suffice to perform a meaningful analysis. Similarly, our fitting procedure uses the full observed frequency range to obtain a singular coupling strength. We investigate how this single value relates to the range of theoretical ones calculated at all frequencies individually. We also test the effect of different treatments of the chemical discontinuity left behind by the first dredge-up.

Knowing the coupling strength is vital for calculating the mixed mode eigenfrequencies of stellar models asymptotically, which is an efficient approach to asteroseismic analysis. Also, the localized link of $q$ to the stellar structure means that it potentially enables the inference of structure parameters from observations. For both applications, it is essential to know which coupling prescription to use in which circumstances. This is the objective of the present study.

\section{Theory}

\subsection{Asymptotic eigenfrequency pattern}

The eigenfrequencies $\nu$ of pure p-modes are to first order evenly spaced \citep{lit:Shibahashi1979}, hence for radial order~$n_\mathrm{p}$ and angular degree~$l$, they are typically written as
\begin{align}
    \nu_{n_\mathrm{p},l}^\mathrm{p}=\Delta\nu\cdot\left(n_\mathrm{p}+\epsilon_{\mathrm{p},l}\right)\,,
    \label{eq:acoustic_resonance}
\end{align}
with the large frequency separation~$\Delta\nu$ (proportional to the inverse of the sound travel time through the star) and p-mode offset~$\epsilon_{\mathrm{p},l}$ induced by the behavior at the turning points \citep{lit:Tassoul1980}. Gravity modes, on the other hand, are asymptotically equidistant in period space \citep{lit:Shibahashi1979}. Therefore, the eigenfrequency of a g-mode of radial order~$n_\mathrm{g}$ ($n_\mathrm{g}<0$ by convention) and degree~$l$ is usually expressed in terms of the oscillation period~$\Pi=\nu^{-1}$ as
\begin{align}
    \Pi_{n_\mathrm{g},l}^\mathrm{g}=\Delta\Pi_l\cdot\left(-n_\mathrm{g}+\epsilon_\mathrm{g}+\frac{1}{2}\right)\,,
    \label{eq:buoyancy_resonance}
\end{align}
with the g-mode offset~$\epsilon_\mathrm{g}$ and the period spacing~$\Delta\Pi_l$, which is inversely proportional to the integral of the Brunt-Väisälä frequency over the g-mode cavity \citep{lit:Tassoul1980}.

By the exchange of mode energy through the evanescent zone, oscillations of the same angular degree~$l$ in the two cavities can couple and form mixed modes. The loss and gain of mode energy at the turning points imposes a shift relative to the eigenfrequencies of the pure modes, such that the resonance condition for mixed modes can be formulated as \citep{lit:Takata2016b}
\begin{align}
    \tan(\Theta_\mathrm{p})\cot(\Theta_\mathrm{g})=q\,,
    \label{eq:mixed_resonance}
\end{align}
where a coupling strength $q=0$ means no and $q=1$ means maximum coupling. The phases~$\Theta_\mathrm{p,g}$ describe the deviation of the mixed mode frequency from the corresponding pure p- and g-mode, respectively (e.g.\,\citet{lit:Shibahashi1979,lit:Takata2016b}; see also \citet{lit:Jiang2018}),
\begin{align}
    \Theta_\mathrm{p}&=\pi\left(\frac{\nu}{\Delta\nu}-(n_\mathrm{p}+\epsilon_\mathrm{p})\right)\,,\label{eq:theta_p}\\
    \Theta_\mathrm{g}&=\pi\left(\frac{\Pi}{\Delta\Pi_l}-(-n_\mathrm{g}+\epsilon_\mathrm{g})\right)\,.\label{eq:theta_g}
\end{align}
Since radial g-modes do not exist and the observation of mixed modes with degree $l\geq2$ is difficult due to a weaker coupling (caused by a smaller extent of the p-mode cavity and hence wider evanescent zone), we focus on the study of dipole modes in this paper. We therefore drop the index $l$ for degree-dependent quantities whenever referring to $l=1$ in the following.

\subsection{Link to stellar structure}

Using a propagating-wave ansatz similar to that used to study the tunneling of quantum-mechanical wave functions through a potential barrier, \citet{lit:Takata2016b} derived a general formula that links the coupling strength to the more fundamental transmission coefficient $T$, the square of which describes the fraction of mode energy passing through the evanescent zone rather than being reflected back into the cavity:
\begin{align}
    q=\frac{1-\sqrt{1-T^2}}{1+\sqrt{1-T^2}}\,.
    \label{eq:q_of_T}
\end{align}
The coefficient $T$ can be linked to stellar structure parameters in the evanescent zone. In this way, the study of mixed modes provides a direct window to the region around the bottom of the convective envelope. So far, this has been discussed in two different limits: the weak and the strong coupling approximation.

\begin{figure}
    \resizebox{\hsize}{!}{\includegraphics{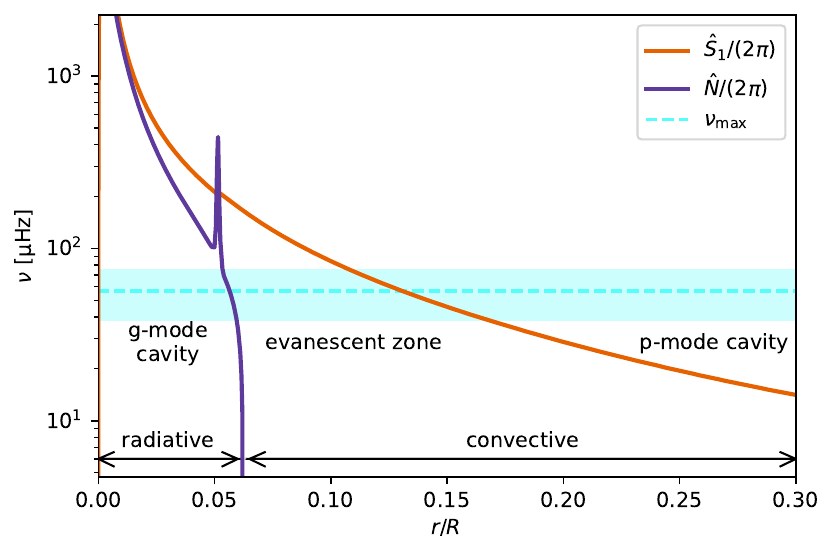}}
    \caption{Propagation diagram of the inner $30\%$ of a red-giant model with $M=1.25\,\mathrm{M}_\odot,\,Z=0.020$: The characteristic frequencies $\hat{S}/(2\pi)$ (orange) and $\hat{N}/(2\pi)$ (purple) as a function of fractional radius. The blue dashed line and shaded area show the value of $\nu_\mathrm{max}$ and the range of dipole mode frequencies used in the analysis. In the g-mode cavity of a mode its frequency lies below, in the p-mode cavity above both characteristic frequencies. Where $\hat{N}/(2\pi)<\nu<\hat{S}/(2\pi)$, the mode evanesces. The chemical discontinuity caused by the first dredge-up (i.e.\,the spike in $\hat{N}/(2\pi)$) is here located just below the evanescent zone of the maximum sampled frequency.}
    \label{fig:propagation_diagram}
\end{figure}

In the weak coupling approximation \citep[first described by][]{lit:Shibahashi1979}, the evanescent zone is assumed to be wide, such that the turning points can be neglected and the behavior of perturbations to the stellar structure can be described by the asymptotic dispersion relation
\begin{align}
    k^2=\frac{\omega^2}{c^2}\left(\frac{\hat{S}^2}{\omega^2}-1\right)\left(\frac{\hat{N}^2}{\omega^2}-1\right)\,,
    \label{eq:dispersion_relation}
\end{align}
with angular frequency $\omega=2\pi\nu$, local sound speed $c$ and the characteristic frequencies $\hat{S}$ (Lamb frequency) and $\hat{N}$ (Brunt-Väisälä frequency), the behavior of which across the stellar profile is shown in Fig.\,\ref{fig:propagation_diagram}. The hat indicates that these frequencies are modified from their canonical form, in order to take perturbations to the gravitational potential into account. \citet{lit:Takata2006b} showed that this modification can be reduced to a factor:
\begin{align}
    \hat{S}&=JS\,,\quad\hat{N}=\frac{N}{J}\,,\label{eq:modified_frequencies}\\
    J&=1-\frac{\rho}{\rho_\mathrm{in}}\,,\label{eq:J_factor}
\end{align}
where $\rho$ is the local density of stellar material and $\rho_\mathrm{in}$ the average density of the spherical volume within the radius at which it is evaluated. The characteristic frequencies $S$ and $N$ are functions of fundamental stellar structure parameters such as density $\rho$, pressure $p$, gravitational acceleration $g$, local sound speed $c$, adiabatic coefficient $\Gamma_1$, and gradients with respect to the radius coordinate $r$ thereof. In oscillation cavities, $k^2>0$ denotes the square of the radial wave vector; conversely, in evanescent zones, $k^2<0$ corresponds to the squared exponential decay length of the general radial solution $x(r)\propto \mathrm{e}^{\mathrm{i}kr}$. The transmission coefficient in the weak coupling approximation $T_\mathrm{w}$ is therefore given by the e-foldings across the evanescent zone:
\begin{align}
    T_\mathrm{w}=\exp\left(\mathrm{i}\int\limits_{r_\mathrm{in}}^{r_\mathrm{out}}k\,\mathrm{d}r\right)=\exp\left(-\int\limits_{r_\mathrm{in}}^{r_\mathrm{out}}\sqrt{|k^2|}\,\mathrm{d}r\right)\,,
    \label{eq:T_weak}
\end{align}
where the limits of the integral are the inner ($r_\mathrm{in}$) and outer ($r_\mathrm{out}$) boundaries of the evanescent zone (i.e.\,zeros of $k^2$). Applying a Taylor expansion to Eq.\,(\ref{eq:q_of_T}) (since $T_\mathrm{w}$ is expected to be small) and inserting Eq.\,(\ref{eq:T_weak}) for $T$ gives the expression found by \citet{lit:Shibahashi1979} (although via different reasoning) for the coupling strength in the weak coupling limit
\begin{align}
    q_\mathrm{w}=\frac{T_\mathrm{w}^2}{4}=\frac{1}{4}\exp\left(-2\int\limits_{r_\mathrm{in}}^{r_\mathrm{out}}\sqrt{|k^2|}\,\mathrm{d}r\right)\,,
    \label{eq:q_weak}
\end{align}
which we will refer to as ``weak coupling'' in the following.

\citet{lit:Takata2016a} derived the transmission coefficient in the strong coupling approximation. The treatment of the singularities in the oscillation equations at $r_\mathrm{in,out}$ introduces an additional term to the transmission coefficient in this limit,
\begin{align}
    T_\mathrm{s}=\exp\left(-\int\limits_{r_\mathrm{in}}^{r_\mathrm{out}}\sqrt{|k^2|}\,\mathrm{d}r-\nabla^2(r_0)\right)\,,
    \label{eq:T_strong}
\end{align}
which we here abbreviate as $\nabla^2(r_0)$ to indicate that it is a positive term containing gradients of the characteristic frequencies, which need to be evaluated at a radius $r_0$ located in the middle of the evanescent zone. $\nabla^2(r_0)\geq0$ implies that $T_\mathrm{s}\leq T_\mathrm{w}$. Physically, this means that the behavior of oscillations around the turning points suppresses the coupling further from the exponential decay over the width of the evanescent zone. However, this expression was derived explicitly assuming a narrow evanescent zone such that both turning points affect the form of the wave function simultaneously, and the contribution from $\nabla^2(r_0)$ is comparable with that from the decay length, while this effect is overestimated if the evanescent zone becomes too wide. For the full expression for $\nabla^2$ as a function of radius, see Appendix \ref{ap:grad_Takata}. Inserting Eq.\,(\ref{eq:T_strong}) into the general expression Eq.\,(\ref{eq:q_of_T}) gives what we will call the ``strong coupling'',
\begin{align}
    q_\mathrm{s}=\frac{1-\sqrt{1-T_\mathrm{s}^2}}{1+\sqrt{1-T_\mathrm{s}^2}}\,.
    \label{eq:q_strong}
\end{align}

Since both expressions have been derived in the framework of asymptotic analysis, they share the underlying assumption that the stellar structure (in particular, the characteristic frequencies) vary only slowly across the evanescent zone compared to the local decay length. However, since $T_\mathrm{s}$ contains gradients of $\hat{N}$ and $\hat{S}$, the value of the strong coupling will be more severely affected by the violation of this assumption.

\section{Methods}\label{s:methods}

\subsection{Stellar models}\label{ss:models}

\begin{figure}
    \resizebox{\hsize}{!}{\includegraphics{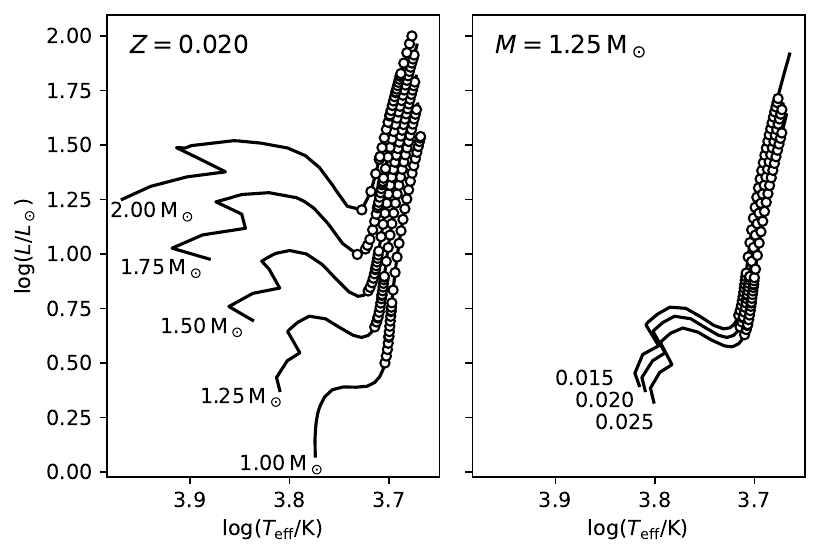}}
    \caption{Evolutionary tracks of stars with various masses $M$ at initial metallicity $Z=0.020$ (left) and various $Z$ at $M=1.25\,\mathrm{M}_\odot$ (right). Circles mark the positions of the models for which we evaluated the coupling strength.}
    \label{fig:HRD_selections}
\end{figure}

We calculated multiple models on the RGB along 7~tracks with masses\footnote{Since mass loss is negligible during the evolutionary phases up to the most evolved model we study, we did not consider it in our models.}
$M\in\{1.00,1.25,1.50,1.75,2.00\}\,\mathrm{M}_\odot$ (for $Z=0.02$) and initial metallicities $Z\in\{0.015,0.020,0.025\}$ (for $M=1.25\,\mathrm{M}_\odot$) using version r24.08.1 of the publicly available stellar evolution code MESA \citep[``Modules for Experiments in Stellar Astrophysics'';][]{lit:MESA1,lit:MESA2,lit:MESA3,lit:MESA4,lit:MESA5,lit:MESA6}. The input physics are standard, including the solar composition described by \citet{lit:Grevesse1998}, and rotation, overshooting and magnetic fields are neglected. However, we added a slight artificial diffusion to smooth the profiles, which improves the calculation of the strong coupling (cf.\,Sect.\,\ref{ss:calculation}). We made sure that this did not significantly affect the evolution of the stars.\footnote{Our MESA inlists and \texttt{run\_star\_extras.f90} are publicly available on Zenodo (doi:~10.5281/zenodo.15261756).} The output of the models notably contains values for $\rho,p,g,\Gamma_1,c,S,N$ and the polytropic index $\frac{\mathrm{d}\ln p}{\mathrm{d}\ln\rho}$ along the radial profile of the star, as well as global parameters like luminosity $L$, effective temperature $T_\mathrm{eff}$, and an estimate for $\nu_\mathrm{max}$ from the scaling relation Eq.\,(\ref{eq:numax_scaling}). In Fig.\,\ref{fig:HRD_selections}, we show the computed tracks in Hertzsprung-Russel diagrams.

The profile data serve as input to the stellar oscillation code GYRE \citep{lit:GYRE}, which we used to solve the oscillation equations numerically and compute the order, degree, and eigenfrequency of modes the star can resonate at. For each model, we calculated the radial ($l=0$) and dipole ($l=1$) modes within $[\nu_\mathrm{max}-3\Delta\nu,\nu_\mathrm{max}+3\Delta\nu]$, with an estimate for $\Delta\nu$ from the sound crossing time. Since the numeric solution does not rely on asymptotic approximations, we can regard it as an independent way to obtain oscillation frequencies and will refer to them as ``observed'' frequencies.

\subsection{Coupling strength calculation}\label{ss:calculation}

With the characteristic frequencies, sound speed, and density as a function of radius given from the models, we calculated the squared asymptotic radial wave vector $k^2$ as per the dispersion relation Eq.\,(\ref{eq:dispersion_relation}) and determined its zeros via sign changes. Discarding zeros close to the stellar center or surface (and potentially those from the spike in the Brunt-Väisälä frequency, i.e.\,the sharp feature in $\hat{N}/(2\pi)$ in Fig.\,\ref{fig:propagation_diagram} arising from the mean-molecular-weight discontinuity caused by the first dredge-up), we identified the remaining two as the limits of the evanescent zone. The integral in Eqs.\,(\ref{eq:q_weak}, \ref{eq:T_strong}) can then be approximated by a Riemann sum between these turning points.

The calculation of the gradient term $\nabla^2$ as a function of the radius coordinate is outlined in Appendix~\ref{ap:grad_Takata}. With the turning points identified as above, we were also able to calculate $r_0=\sqrt{r_\mathrm{in}r_\mathrm{out}}$ according to Eq.\,(\ref{eq:definition_s}). We then used the two grid points in the stellar profile with radius coordinates closest to $r_0$ to estimate $\nabla^2(r_0)$ by linear interpolation between them. When the spike in the Brunt-Väisälä frequency lies in the evanescent zone, the steep variation of the characteristic frequency and the additional zeros in $k^2$ violate the underlying assumption of smooth variations in the stellar structure and make it impossible to evaluate this term meaningfully (see \citet{lit:Jiang2022b} for a detailed analysis of the effects) and hence we did not attempt the calculation of $q_\mathrm{s}$ for frequencies where this is the case.

With the integral and gradient terms evaluated in these ways, we calculated both, $q_\mathrm{w}$ and $q_\mathrm{s}$, for each model, for the corresponding $\nu_\mathrm{max}$ as well as for all dipole mode frequencies identified by GYRE that we used in the asymptotic fit described in the following section. In this way, we probed the frequency dependence of $k^2$ and its zeros (cf.\,Eq.\,(\ref{eq:dispersion_relation})).

To investigate the effect of the Brunt-Väisälä-spike on the weak coupling and the evolution of the strong coupling without this feature, we evaluated both couplings a second time for each model after removing the spike by linearly interpolating $N^2$ and the polytropic index between the spike's base-points.

\subsection{Asymptotic fitting}\label{ss:fitting}

\begin{figure}[t]
    \resizebox{\hsize}{!}{\includegraphics{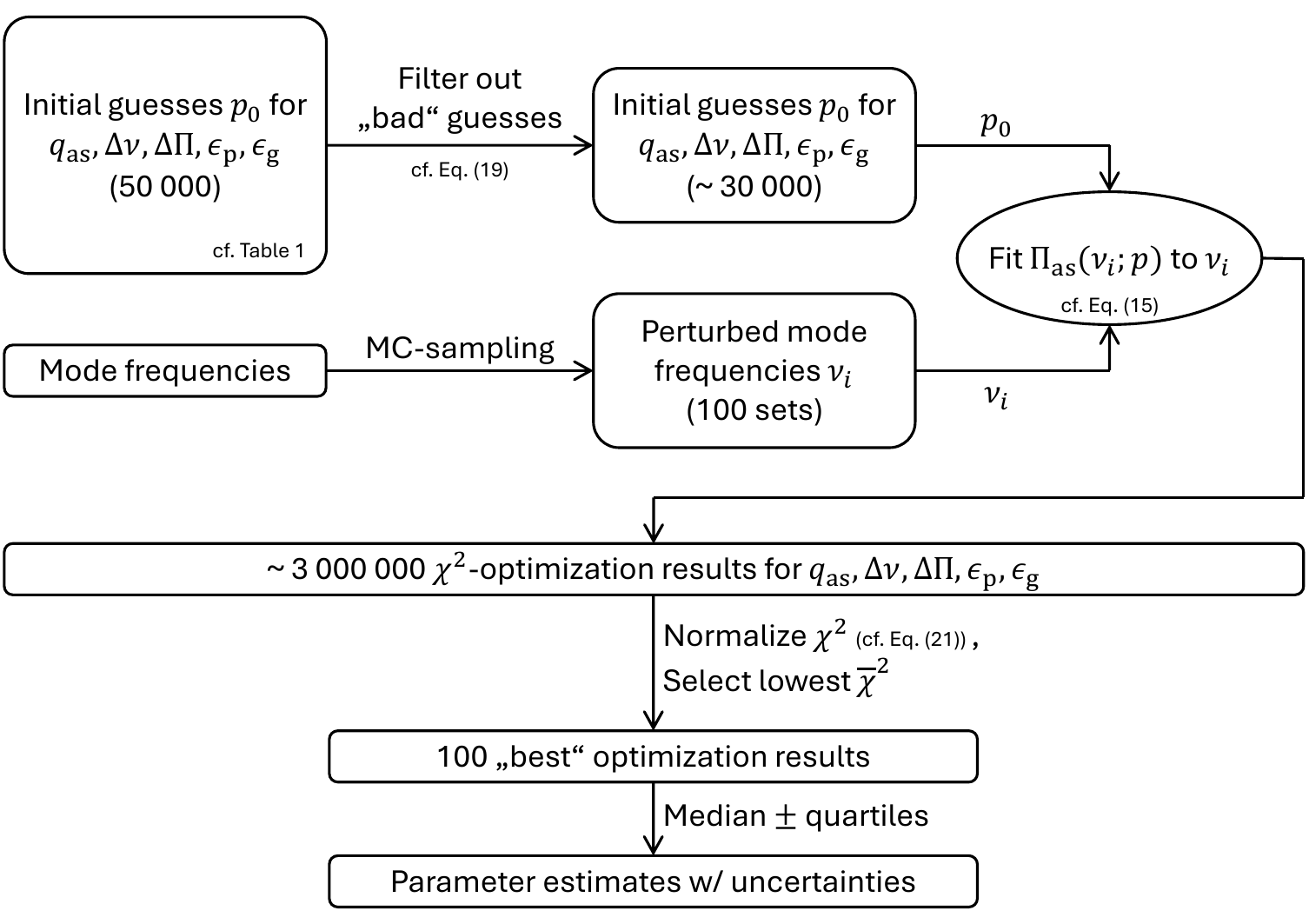}}
    \caption{Flowchart outlining our fitting procedure.}
    \label{fig:flowchart}
\end{figure}

Starting from the mixed mode resonance condition Eq.\,(\ref{eq:mixed_resonance}), an asymptotic formula for the period of mixed modes can be derived following the arguments by \citet{lit:CD2012}, which has been used in a similar way by \citet{lit:Hekker2018}. Inserting the definitions Eqs.\,(\ref{eq:theta_p},~\ref{eq:theta_g}) into Eq.\,(\ref{eq:mixed_resonance}) and rearranging the terms gives
\begin{align}
    \Pi_\mathrm{as}=\Delta\Pi\cdot\left(-n_\mathrm{pg}+\epsilon_\mathrm{g}+\frac{1}{2}-\frac{\arctan\left[q\cot\left(\pi\left(\frac{\nu}{\Delta\nu}-\epsilon_\mathrm{p}\right)\right)\right]}{\pi}+n_\mathrm{p}\right)\,,
    \label{eq:Pi_as}
\end{align}
where the periodicity of the tangent function with $m\pi,\,m\in\mathbb{Z}$ is used to replace $n_\mathrm{g}$ by the more straightforward total radial order $n_\mathrm{pg}=n_\mathrm{p}+n_\mathrm{g}$, which always increases by $1$ when going to the mode with same degree and next higher eigenfrequency (given that all consecutive modes are present in the sample). Adding the acoustic radial order $n_\mathrm{p}$ lifts the degeneracy introduced when inverting the tangent of the expression in such a way that $\Pi_\mathrm{as}$ becomes a smooth function of $\nu$. Since $n_\mathrm{p}$ is sometimes mis-assigned by GYRE, we replaced it by a parameter $n_\varphi$ which explicitly fulfills this purpose, by defining
\begin{align}
    n_{\varphi,0}&=n_{\mathrm{p},0}\,,\\
    n_{\varphi,i\,\geq1}&=\left\{\begin{array}{ll}n_{\varphi,i-1}\,,&\varphi_i\leq\varphi_{i-1}\\n_{\varphi,i-1}+1\,,&\varphi_i>\varphi_{i-1}\end{array}\right.\,,
    \label{eq:nPhi}
\end{align}
where the index $i$ labels the observed modes in order of ascending eigenfrequency and $\varphi=\arctan\left[q\cot\left(\pi\left(\frac{\nu}{\Delta\nu}-\epsilon_\mathrm{p}\right)\right)\right]$.

In observations, the radial orders $n_\mathrm{pg}$ and $n_{\mathrm{p},0}$ would be unknown and hence appear as free parameters. Working with models, however, we were able to treat them as given quantities, just as the eigenfrequencies. The five remaining parameters are $\{q_\mathrm{as},\Delta\nu,\Delta\Pi,\epsilon_\mathrm{p},\epsilon_\mathrm{g}\}$, where we will refer to the coupling strength obtained from the asymptotic periods as ``asymptotic coupling''. We fitted for the five free parameters by minimization of the $\chi^2$-sum
\begin{align}
    \chi^2=\sum\limits_i\left[\frac{1}{\nu_i}-\Pi_\mathrm{as}(\nu_i,n_{\mathrm{pg},i},n_{\varphi,i};q_\mathrm{as},\Delta\nu,\Delta\Pi,\epsilon_\mathrm{p},\epsilon_\mathrm{g})\right]^2
    \label{eq:chisq}
\end{align}
over the modes $i$ we found for each model.

This $\chi^2$-sum has multiple local minima. Also, the parameters are strongly correlated. Both these facts make numerical optimization difficult, since the minimization algorithm can converge to a local rather than the global minimum in one parameter, which will then also affect the values of the other parameters. To circumvent this, we performed multiple optimizations per model with varying initializations. The flowchart in Fig.\,\ref{fig:flowchart} visually represents the fitting procedure outlined in the following paragraphs.

\begin{table}[t]
\caption{Properties of the initial guess sampling}
\centering
\begin{tabular}{llll}
\hline\hline Parameter&Estimate&Width&Distribution\\\hline
$q_\mathrm{as}$ & 0.2 & 0.2 & Gaussian \\
$\Delta\nu$ & $\Delta\nu_0$ ($l=0$-fit) & $0.03\cdot\Delta\nu_0$ & Gaussian \\
$\Delta\Pi$ & $\Delta\Pi_N$ ($N$-integral) & $0.03\cdot\Delta\Pi_N$ & Gaussian \\
$\epsilon_\mathrm{p}$ & $\epsilon_{\mathrm{p},0}+\frac{1}{2}$ & 0.7 & uniform \\
& ($\epsilon_{\mathrm{p},0}$: $l=0$-fit) & & \\
$\epsilon_\mathrm{g}$ & 0.5 & 1.2 & uniform\\\hline
\end{tabular}
\title{}
\tablefoot{For Gaussian distributions, ``width'' refers to the standard deviation of the Gaussian; for uniform distributions the limits are set to estimate$\pm$width. Since $q$ can physically only take positive values, all instances where the Gaussian sampling returns a value $<0$ are resampled with a uniform distribution over $[0,0.8]$.}
\label{tab:parameter_shooting}
\end{table}

From the distributions specified in Table~\ref{tab:parameter_shooting}, we drew 50\,000 sets of initial guesses for the parameters. To save computation time, we only kept those which corresponded to a sufficiently low initial $\chi^2$-sum,
\begin{align}
    \chi^2\leq N_\mathrm{modes}\cdot\left(\frac{0.03}{\langle\nu\rangle}\right)^2\,,
\end{align}
i.e.\,the average deviation is no larger than $3\%$ of the period corresponding to the mean frequency of the sample $\langle\nu\rangle$ -- typically, these were still around 30\,000 sets. Additionally, to avoid chance alignments and simulate observational noise, we applied a Monte Carlo sampling with standard deviation $0.01\,\text{\textmu}\mathrm{Hz}$ and 100~realizations to the observed frequencies. For each combination of initial parameter guesses and perturbed frequencies, we then used the \texttt{curve\_fit} function included in the python-package \texttt{scipy.optimize} with the implementation of the ``dogbox'' algorithm to minimize the $\chi^2$-sum Eq.\,(\ref{eq:chisq}). To ensure that all local minima were sampled, we only allowed the parameters $\Delta\nu$ and $\Delta\Pi$ -- which are most strongly subjected to aliasing -- to vary by $\pm0.5\%$ from the initial guess. In turn, we set the tolerance for the termination of the algorithm rather high, which has the side effect that the fit sometimes converges ``too early'' close to the limits. This especially affects the fitting for $q_\mathrm{as}$, which has a hard limit at 0. In models where the actual value of $q$ is low (less than $\sim0.1$), the $\chi^2$-sum of these limit convergences can become comparable to the global minimum and therefore affect the statistics of the best-fit distributions. To avoid errors from this effect, we only accepted fits that converged to $q_\mathrm{as}\geq5\cdot10^{-3}$ for further consideration.

We subsequently normalized the optimized $\chi^2$-sum to make it more comparable among the different models. As an estimate for the degrees of freedom, we used the number of modes available to the fit $N_\mathrm{modes}$ minus the number of expected modes per acoustic order, which define the shape of the mixed mode pattern and therefore need to be fixed while the availability of multiple acoustic orders introduces the redundancy allowing for the fit to improve. Since the width of an acoustic order is given by $\Delta\nu$, we approximated the latter by
\begin{align}
    N_{\Delta\nu}=\frac{\Delta\nu}{\Delta\Pi\cdot\langle\nu\rangle^2}\,,
    \label{eq:N_per_order}
\end{align}
to give the normalized $\chi^2$-sum
\begin{align}
    \overline{\chi}^2=\frac{\chi^2}{N_\mathrm{modes}-N_{\Delta\nu}}\,.
    \label{eq:chisq_norm}
\end{align}
This normalization would also disfavor aliases arising from oversampling of the modes if the $n_\mathrm{pg}$ were unknown, because the higher chance for some of the asymptotic modes to align well with the observed ones at a higher density of mixed modes would be counteracted by the increase in $N_{\Delta\nu}$. By this normalized $\overline{\chi}^2$-sum, we selected the 100 best fits (i.e.\,those with lowest $\overline{\chi}^2$) and calculated the final estimate of the fit parameters as the median of their distribution among this ``best'' subset, with uncertainties as the difference to the first and third quartile, respectively.

\section{Results}\label{s:results}

\begin{figure*}
    \centering
    \includegraphics[width=17cm]{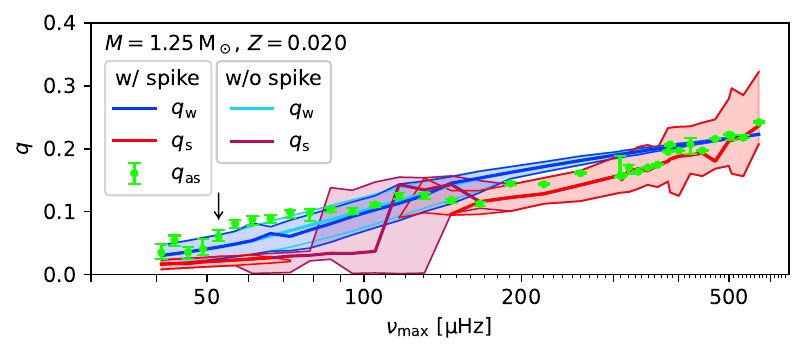}
    \caption{Values of the coupling strength $q$ computed in the three ways presented in Sect.\,\ref{s:methods} as a function of $\nu_\mathrm{max}$ for stellar models with $M=1.25\,\mathrm{M}_\odot,\,Z=0.020$ along the RGB. The star evolves from right to left. Solid lines represent $q_\mathrm{w}(\nu_\mathrm{max})$ (dark blue for the full profile; light blue for profile with the $N^2$-spike removed) and $q_\mathrm{s}(\nu_\mathrm{max})$ (red for full profile, purple without spike), respectively, shaded areas show the range of values calculated for the frequencies in the sample used for the fit; green markers show $q_\mathrm{as}$ with uncertainties as described in Sect.\,\ref{ss:fitting}. Below $100\,\text{\textmu Hz}$, the x-axis ticks show spacings of $10\,\text{\textmu Hz}$; above $100\,\text{\textmu Hz}$, major ticks show spacings of $100\,\text{\textmu Hz}$ with minor ticks continuing in steps of $10\,\text{\textmu Hz}$. The arrow indicates the model further analyzed in Fig.\,\ref{fig:q_as_of_nu}.}
    \label{fig:q_of_numax}
\end{figure*}

In Fig.\,\ref{fig:q_of_numax}, we show the results of the computations outlined in Sect.\,\ref{s:methods} for a modeled star with $M=1.25\,\mathrm{M}_\odot$ and $Z=0.02$ along its evolution up the RGB. The coupling was evaluated for the models indicated with circles in the Hertzsprung-Russel diagram in Fig.\,\ref{fig:HRD_selections}. Since a star ascending the RGB expands in radius at (nearly) constant mass, $\nu_\mathrm{max}$ decreases with evolution, meaning that the star moves from right to left in the figure. We visually analyze trends in this and similar figures to discuss our results.

\subsection{Applicability of the approximations}\label{ss:applicability}

Initially, the asymptotic coupling is in agreement with the values predicted in the strong coupling approximation. Where the ranges of $q_\mathrm{w}$ and $q_\mathrm{s}$ overlap, $q_\mathrm{as}$ is in some cases also consistent with the weak coupling approximation. Coming from large values on the sub-giant branch, the data show that the asymptotic coupling follows $q_\mathrm{s}$ along the early RGB (for $\nu_\mathrm{max}\ga110\,\text{\textmu Hz}$), even when the reduction of transmitted mode energy due to the second term in Eq.\,(\ref{eq:T_strong}) is so strong that $q_\mathrm{s}<q_\mathrm{w}$. In more evolved models ($\nu_\mathrm{max}\la70\,\text{\textmu Hz}$), the strong coupling has very low values. The data points show, however, that in this regime the weak coupling approximation is more appropriate to describe the coupling observed in the mixed mode pattern.

In intermediate models ($110\,\text{\textmu Hz}\ga\nu_\mathrm{max}\ga70\,\text{\textmu Hz}$), $q_\mathrm{s}$ could not be calculated from the full stellar profile due to the presence of the spike in the Brunt-Väisälä frequency (as discussed in Sect.\,\ref{ss:calculation}). Removing the spike reveals that the strong coupling drops to the very low values quite steeply. The timestep at which this drop occurs is frequency-dependent, which explains the wide range of values $q_\mathrm{s}$ takes in this regime as represented by the purple shaded area.

We find that the weak coupling gives a good estimate for the asymptotic values in this intermediate regime. The comparison of $q_\mathrm{w}$ for profiles with and without the spike shows that the spike mainly increases the integral over the decay length $\sqrt{|k^2|}$ and hence gradually decreases the coupling as it moves up through the evanescent zone. When $\nu_\mathrm{max}$ falls below $\hat{N}^2/(2\pi)$ at the base of the spike, the evanescent zone abruptly narrows and $q_\mathrm{w}$ steps up to agree with the spike-free value again. However, we see that the deviation is marginal compared to the range given by the frequency dependence.

When the asymptotic coupling is consistent with the strong coupling approximation, the fitted values mostly fall close to the middle of the range in coupling strengths $q_\mathrm{s}$ evaluated at the different frequencies, while they tend to lie closer to the upper end of the shaded area when they follow the weak coupling.

\subsection{Mass- and metallicity-dependence}\label{ss:MZ-dependence}

\begin{figure}
    \resizebox{\hsize}{!}{\includegraphics{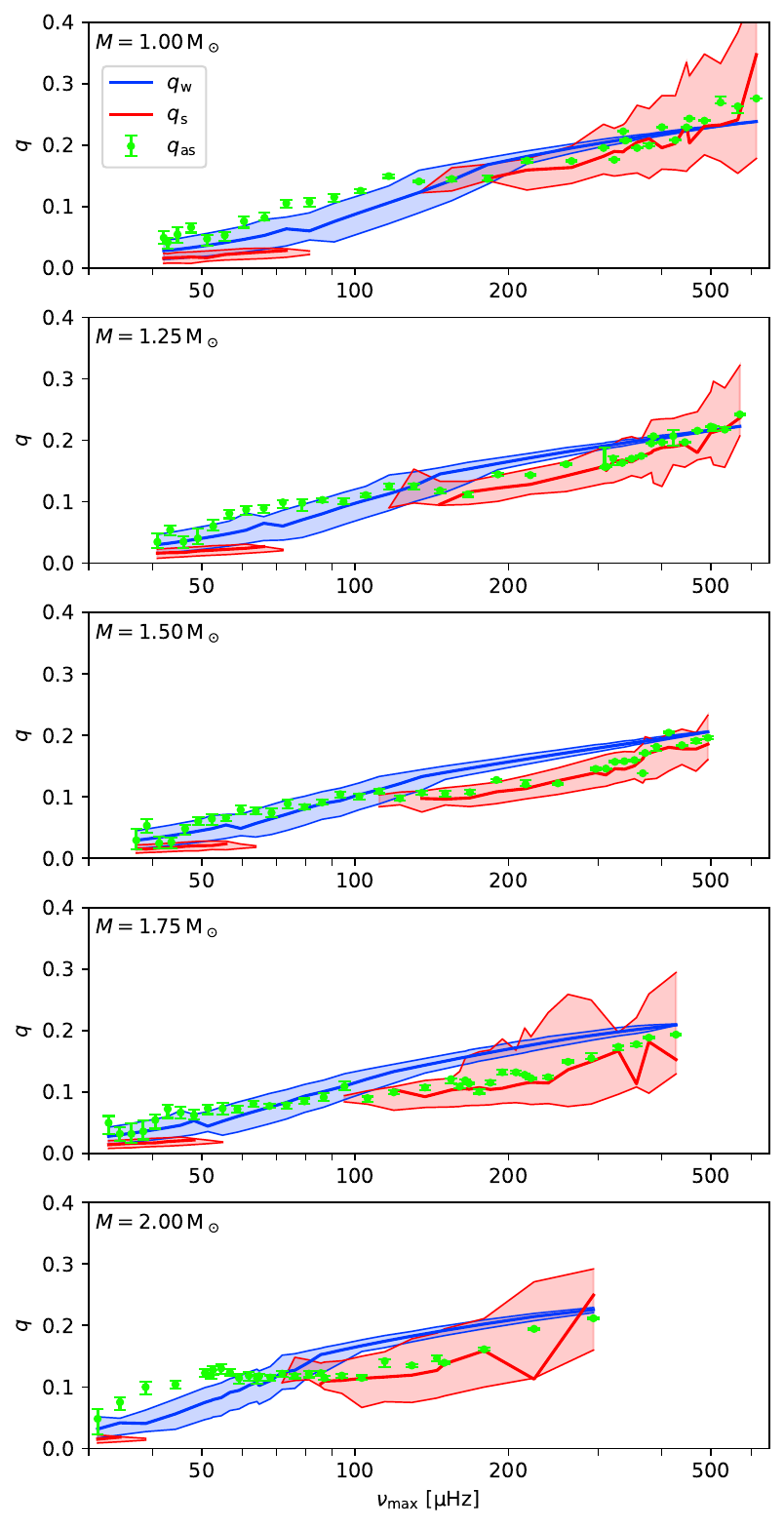}}
    \caption{Values of the coupling strength $q$ as a function of frequency of maximum oscillation power $\nu_\mathrm{max}$ for RGB models with different masses $M$ as indicated in each panel and initial metallicity $Z=0.020$. See~Fig.\,\ref{fig:q_of_numax} for the meaning of colors and symbols. Stars evolve from right to left.}
    \label{fig:q_of_numax_M}
\end{figure}

\begin{figure}
	\resizebox{\hsize}{!}{\includegraphics{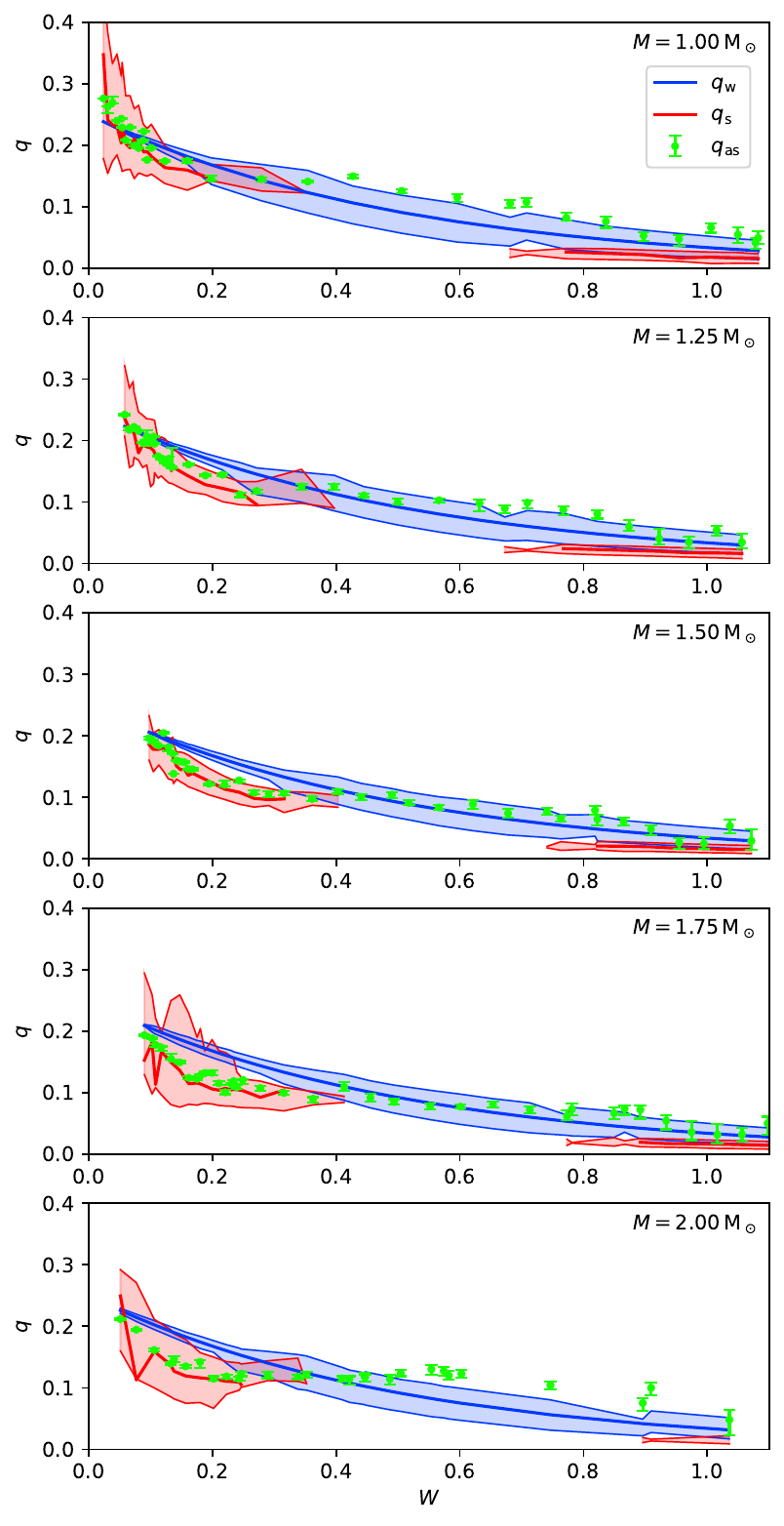}}
	\caption{Values of the coupling strength $q$ as a function of evanescent zone width $W$ for RGB models with different masses $M$ as indicated in each panel and initial metallicity $Z=0.020$. See~Fig.\,\ref{fig:q_of_numax} for the meaning of colors and symbols. Stars evolve from left to right.}
	\label{fig:q_of_kr_M}
\end{figure}

While the mass of the red-giant affects the value of the coupling as a function of $\nu_\mathrm{max}$ (cf.\,Fig.\,\ref{fig:q_of_numax_M}), the overall trend stays the same: Coming from the sub-giant branch, where the coupling can reach very high values, $q_\mathrm{as}$ continues to closely follow the strong coupling approximation as long as the calculation in this limit is still possible. Also, the models with the spike removed show a similar behavior relative to the ones with the spike included, so the results of the former are omitted for clarity.

With increasing mass, the onset of the RGB phase occurs at lower $\nu_\mathrm{max}$, since the corresponding radius also increases. On the lower RGB, the coupling strength is lower for higher stellar mass at the same $\nu_\mathrm{max}$. To exclude that this effect is solely due to the radius evolution as well, we plot $q$ as a function of the evanescent zone width in terms of the number of asymptotic e-foldings $W$,
\begin{align}
	W\coloneqq\int\limits_{r_\mathrm{in}(\nu_\mathrm{max})}^{r_\mathrm{out}(\nu_\mathrm{max})}\sqrt{|k^2(\nu_\mathrm{max})|}\,\mathrm{d} r\,.
\label{eq:definition_W}
\end{align}
Figure~\ref{fig:q_of_kr_M} shows that indeed for the same width $W$, the coupling tends to be weaker in higher-mass stars on the early RGB (most prominently at $W\sim0.2$). With $q_\mathrm{w}(\nu_\mathrm{max})$ being a mass-independent function of $W$ as per Eq.\,(\ref{eq:q_weak}), the blue line can serve as a good visual reference.

As soon as the calculation of the strong coupling approximation breaks down due to the spike in the Brunt-Väisälä frequency, all models across the mass range considered in this work can be described using the weak coupling approximation within three times the width of the uncertainties. While the $\nu_\mathrm{max}$ at which this transition occurs varies with mass (cf.\,Fig.\,\ref{fig:q_of_numax_M}), the corresponding evanescent zone width is fairly constant at $W\sim0.4$ (cf.\,Fig.\,\ref{fig:q_of_kr_M}).

\begin{figure}
	\resizebox{\hsize}{!}{\includegraphics{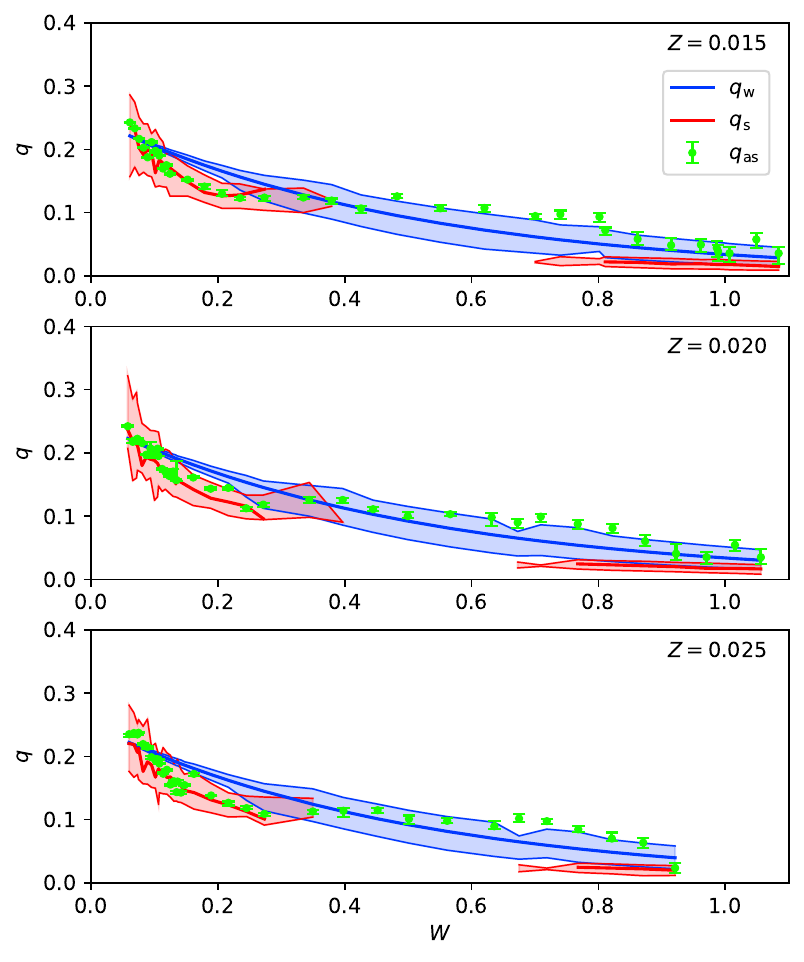}}
	\caption{Values of the coupling strength $q$ as a function of evanescent zone width $W$ for RGB models with different initial metallicities $Z$ as indicated in each panel and mass $M=1.25\,\mathrm{M}_\odot$. See~Fig.\,\ref{fig:q_of_numax} for the meaning of colors and symbols. Stars evolve from left to right.}
	\label{fig:q_of_kr_Z}
\end{figure}

Metallicity does not significantly influence the coupling strength of dipolar mixed modes in red-giants. Figure~\ref{fig:q_of_kr_Z} shows that the evolution of $q$ is very similar for the three $1.25\,\mathrm{M}_\odot$-models calculated with different initial metallicities.

\section{Discussion}\label{s:discussion}

\begin{figure*}
    \centering
    \includegraphics[width=17cm]{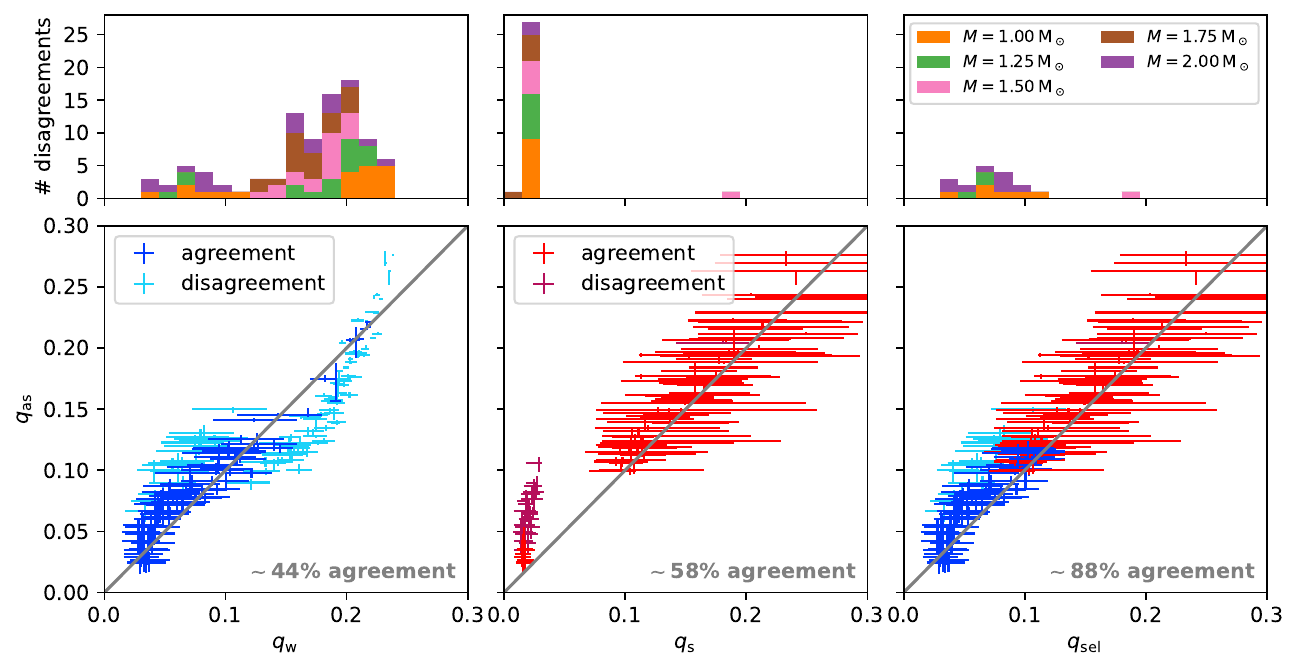}
    \caption{Comparison of coupling strengths obtained via asymptotic fitting and theoretical calculations. From left to right, the theoretical prescription used is the weak coupling, strong coupling, and the selection of one of the two as given by Eq. (\ref{eq:q_sel}).\\
    \textit{Bottom row:} $q_\mathrm{as}$ (with uncertainties as described in Sect.\,\ref{ss:fitting}) on the y-axis versus theoretical values (with error bars indicating range across fitted frequencies) on the x-axis. Models in agreement with the one-to-one line (gray diagonal) within the error bars are shown in dark blue/red for $q_\mathrm{w}$/$q_\mathrm{s}$, those not in agreement in light blue/purple. Percentages relative to total number of models fitted.\\
    \textit{Top row:} Histograms of the distribution of disagreeing models over the theoretical coupling values. Stacked up colors represent models of different masses as indicated in the legend.}
    \label{fig:q_as_vs_q_theo}
\end{figure*}

In agreement with e.g.\,\citet{lit:Pincon2020,lit:Jiang2022a}, we find that along the RGB the coupling strength generally decreases with evolution. While the evanescent zone lies in the radiative layer above the hydrogen-burning shell, the asymptotic coupling is consistent with the values predicted in the strong coupling approximation. Once the evanescent zone is located at the bottom of the convective envelope, $q_\mathrm{as}$ is well described by the weak coupling. Although this distinction by the radiative or convective nature of the evanescent zone is not directly visible from the results presented in Figs. \ref{fig:q_of_numax}--\ref{fig:q_of_kr_Z}, we can infer the two regimes discussed from the transient of the spike in the Brunt-Väisälä frequency (which is visible in the figures from the fact that we cannot calculate $q_\mathrm{s}$ when it lies in the evanescent zone). Since the spike forms at the deepest extent of the convective envelope and the evanescent zone moves outwards during evolution, the latter must lie in the radiative layer before the spike appears. After it has passed through, the evanescent zone is predominantly convective, since the convective envelope has not receded much by this point \citep{lit:vanRossem2024}. This assumption is supported by the analysis of the relevant stellar profiles, for instance Fig.\,\ref{fig:propagation_diagram} for the model with $\nu_\mathrm{max}\sim55\,\text{\textmu Hz}$ in Fig.\,\ref{fig:q_of_numax}. Given that $\hat{N}^2<0$ in convective regions since it is related to the Schwarzschild criterion for convection, this propagation diagram shows that most of the evanescent zone in this model is convective for the considered frequency range.

While \citet{lit:Pincon2020} argued there should be an intermediate regime in which none of the two approximations holds, our model calculations suggest that the range of $q_\mathrm{w}$ is in agreement with $q_\mathrm{as}$ (within at maximum three times the uncertainties) as soon as the width of the evanescent zone $W\geq0.4$ in units of the local decay length. For $W<0.4$, $q_\mathrm{s}$ can still be calculated for at least some frequencies in the observable range. We therefore suggest a selection of the coupling formula to use as described by:
\begin{align}
    q_\mathrm{sel}=\left\{\begin{array}{ll}q_\mathrm{s}\,,&W<0.4\\q_\mathrm{w}\,,&W\geq0.4\end{array}\right..
\label{eq:q_sel}
\end{align}
The quality of this selection for our models is shown in Fig.\,\ref{fig:q_as_vs_q_theo}, which demonstrates the good agreement of the theoretical values with the observed ones using this straightforward prescription. The cutoff of $W=0.4$ corresponds to a weak coupling of $q_\mathrm{w}\sim0.11$ following Eq.\,(\ref{eq:q_weak}).

\begin{figure}
	\resizebox{\hsize}{!}{\includegraphics{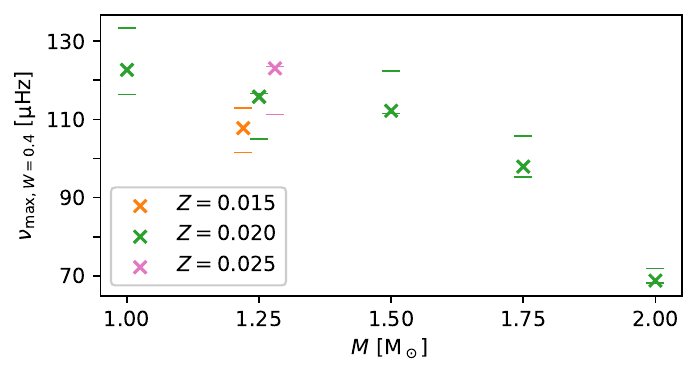}}
	\caption{Values of $\nu_\mathrm{max}$ corresponding to models with threshold evanescent zone width as a function of stellar mass ($M=1.25\,\mathrm{M}_\odot$-markers slightly offset to make separable). Horizontal bars show $\nu_\mathrm{max}$ of the models with $W$ closest to $0.4$ either side, crosses show the linearly interpolated values for $W=0.4$. Metallicity is distinguished by color as indicated in the legend.}
	\label{fig:numax_W_is_04}
\end{figure}

The values of observables such as $\nu_\mathrm{max}$ corresponding to the $W=0.4$-cutoff are mass-dependent: the transition occurs at lower $\nu_\mathrm{max}$ for higher masses (cf.\,Fig.\,\ref{fig:numax_W_is_04}). The dependence is non-linear. Metallicity potentially also affects the value of $\nu_\mathrm{max}$ at which $W=0.4$, albeit less strongly. These dependencies make it difficult to determine whether the evanescent zone of an observed red-giant is radiative or convective. Since the variation of $q$ with evolution is rather slow in this regime, the use of Eq.\,(\ref{eq:q_sel}) to observationally differentiate the two stages is also limited.

\subsection{Strong coupling}

The lower values of $q_\mathrm{s}$ at equal evanescent zone width $W$ for more massive stars can be explained by the fact that, owing to the larger extent of the convective core on the main sequence, a higher proportion of the stellar mass is confined to the inert helium core of the red-giant. As the core contracts and the envelope expands along the RGB, this means that the gradient $\frac{\mathrm{d} g}{\mathrm{d} r}\propto\rho-\frac{2}{3}\rho_\mathrm{in}$ is more strongly negative in the evanescent zone of more massive stars (which lies in the expanding region and hence has low density relative to the core). This increases the gradient of the modified Brunt-Väisälä frequency in Eq.\,(\ref{eq:dlnc_ds}) and thus the second term in Eq.\,(\ref{eq:T_strong}).

Across all masses and metallicities we considered, the width of the red shaded area at small $W\leq0.4$ (i.e.\,on the early RGB) is rather large. This is not due to the real frequency dependence of $q_\mathrm{s}$, but rather to numerical noise along the stellar radial profile that affects the evaluation of the gradient term in Eq.\,(\ref{eq:T_strong}). This is also the reason for the non-monotonous evolution of $q_\mathrm{s}$ with $\nu_\mathrm{max}$ or even $W$. Since a shift in frequency also shifts the turning points $r_\mathrm{in,out}$ and thus $r_0$, where the gradients are evaluated, the centroid of the shaded area can be seen as an estimate for the true strong coupling value, which well reproduces the behavior of $q_\mathrm{as}$ in this regime. This is expected, since the evanescent zone lying in the radiative layer right above the hydrogen-burning shell is very narrow during this phase \citep{lit:vanRossem2024}.

At the bottom of the convective envelope, $\hat{N}^2$ abruptly falls below zero. This drop in the Brunt-Väisälä frequency (while the Lamb-frequency maintains a similar power-law shape to the one it has below the envelope, cf.\,Fig.\,\ref{fig:propagation_diagram}) means that the evanescent zone rapidly widens (i.e.\,$W$ increases), which allows for the weak coupling approximation to become applicable \citep{lit:Pincon2020}. Meanwhile, the second term in Eq.\,(\ref{eq:T_strong}) remains of the same order of magnitude, so $T_\mathrm{s}$ is still strongly suppressed compared to $T_\mathrm{w}$. In fact, the steep gradient of $\hat{N}^2$ around the bottom of the convection zone (which remains the inner turning point $r_\mathrm{in}$) increases $\nabla^2$ everywhere between the turning points, implying that the suppression of $T_\mathrm{s}$ below $T_\mathrm{w}$ is even stronger when the evanescent zone lies in the convective envelope. This explains why the drop of $q_\mathrm{s}$ is quite sudden when the evanescent zone moves into the convective envelope, as shown for the models with the spike removed in Fig.\,\ref{fig:q_of_numax}.

\subsection{Weak coupling}

In models with a convective evanescent zone, the asymptotic coupling strength tends to fall closer to the maximum of $q_\mathrm{w}$ calculated from the frequency sample, rather than to the central value. This can be due to both an error in the weak coupling approximation and an incompleteness in the asymptotic formula~(\ref{eq:Pi_as}), which does not account for the frequency dependence of the coupling \citep[see also][]{lit:Jiang2020}. As discussed in \citet{lit:Pincon2020} (and also shown in Fig.\,\ref{fig:q_of_numax} by the width of the blue shaded area), the frequency dependence of the weak coupling is much stronger for evanescent zones that lie in the convective envelope. This is due to the radial profile of the characteristic frequencies shown in Fig.\,\ref{fig:propagation_diagram}: In the radiative zone, both characteristic frequencies have a similar slope and thus the width of the evanescent zone only changes little when moving to higher or lower frequency. At the bottom of the convective envelope, however, the steep slope of the Brunt-Väisälä frequency keeps the inner turning point almost fixed when changing the frequency. Thus, modes at slightly higher frequencies already have a substantially narrower evanescent zone and are therefore more strongly coupled. Since the asymptotic formula treats $q_\mathrm{as}$ as a frequency-independent parameter, the fit will converge to some ``average'' value.

\begin{figure}
	\resizebox{\hsize}{!}{\includegraphics{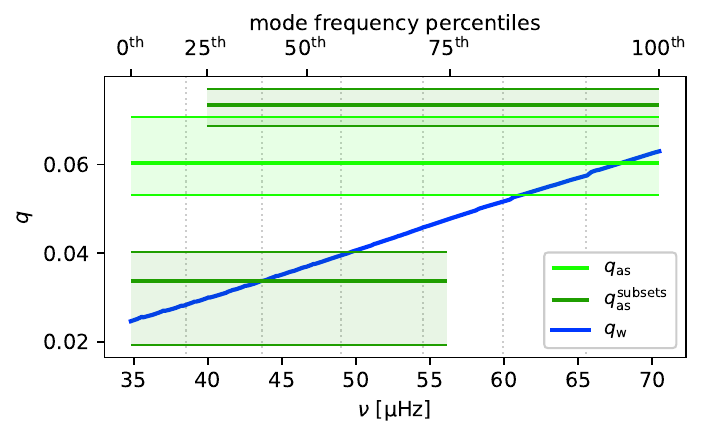}}
	\caption{Coupling strength for the model marked with an arrow in Fig.\,\ref{fig:q_of_numax} as a function of mode frequency. Fitted values $q_\mathrm{as}$ were fit to modes in the frequency range indicated by the width of the marker, uncertainties are estimated as described in Sect.\,\ref{ss:fitting}. The solid dark blue line shows the weak coupling calculated as per Eq.\,(\ref{eq:q_weak}). Vertical dotted lines indicate the frequencies of radial modes.}
	\label{fig:q_as_of_nu}
\end{figure}

To assess whether this ``average'' is biased towards higher frequencies, we performed fits to two subsets of modes: those with frequency below the 75\textsuperscript{th} and above the 25\textsuperscript{th} percentile of the full sample by number of dipole modes, respectively. Fig.\,\ref{fig:q_as_of_nu} shows that indeed the fit over the full frequency range lays significantly closer to that only including the higher frequencies. This suggests that these dominate the $\chi^2$-sum (Eq.\,(\ref{eq:chisq})) when it comes to the variation of $q_\mathrm{as}$, leading to convergence to a parameter set with $q_\mathrm{as}$ systematically higher than $q_\mathrm{w}(\nu_\mathrm{max})$. This can be explained by the fact that at higher frequencies, the number of mixed modes per acoustic order is lower (cf. also Eq.\,(\ref{eq:N_per_order})) and therefore there are more orders (contributing to the ``average'') in a sample with the same number of modes. This suggests that, in order to match observed coupling strengths in the weak coupling regime, it is advisable to calculate $q_\mathrm{w}$ not at $\nu_\mathrm{max}$, but rather at a frequency close to the maximum of the observed range. At the same time, it is clear that the error introduced by the assumptions of a wide evanescent zone (for instance, the Taylor expansion of Eq.\,(\ref{eq:q_of_T}) and the neglect of the behavior of $k$ at the turning points in $T_\mathrm{w}$) affects the more strongly coupled modes with higher frequencies more. The systematic offset of $q_\mathrm{as}$ from $q_\mathrm{w}(\nu_\mathrm{max})$ is therefore a combined effect of the ignored frequency dependence and incompleteness of the weak coupling approximation. For some models with very weak coupling, $q_\mathrm{as}$ falls below $q_\mathrm{w}(\nu_\mathrm{max})$ against this general trend. These might still be subject to the effect of limit convergences, which become very difficult to disentangle from the statistics of the actual best fit subset.

The fact that the weak coupling can be used already for $W\sim0.4$ is somewhat fortuitous, since for a value of $q\sim0.1$, the Taylor expansion used in Eq.\,(\ref{eq:q_weak}) alone deviates by $\sim25\%$ from the full expression; underestimating $q_\mathrm{w}$ for a given transmission coefficient $T_\mathrm{w}$. It appears, however, that this roughly cancels with the fact that neglecting the turning points compared to the derivation by \citet{lit:Takata2016a} overestimates the transmission coefficient in such narrow evanescent zones. This empiric result simplifies the calculation of coupling strengths enormously. We note that the error of this result is mass-dependent: Using the weak coupling approximation directly above $W=0.4$ is least appropriate for $M=2.00\,\mathrm{M}_\odot$ (cf. Fig.\,\ref{fig:q_of_kr_M}), where we have a systematic deviation from the highest $q_\mathrm{w}$ of up to three times the uncertainty in $q_\mathrm{as}$ for models with $0.6\la W\la0.9$. Therefore, we cannot expect to be able to apply this result to stars with even higher masses. The cancellation of the aforementioned effects works best for the tracks with masses of $1.50\,\mathrm{M}_\odot$ and $1.75\,\mathrm{M}_\odot$ (cf.\,top row of Fig.\,\ref{fig:q_as_vs_q_theo}). While the ``actual'' effect of the turning points is not trivial to quantify (which is why only the coupling prescriptions in the limits of explicitly wide and narrow evanescent zones have been derived so far), it is clear that it depends on the stellar structure in a different way than the quality of the Taylor expansion, which is determined by $W$ alone. This different dependence allows the quality of the cancellation to vary with stellar mass in the way we observe it: The error introduced by the Taylor expansion is larger than that coming from the neglect of the turning points in the models with $M\in\{1.00,2.00\}\,\mathrm{M}_\odot$, while they are similar in models with masses in between.

In principle, while the spike at the mean molecular weight discontinuity lies in the evanescent zone, it should act as another buoyancy cavity where mode energy can be trapped; however, in the practical calculation of $q_\mathrm{w}$, we ignore it and integrate $\sqrt{|k^2|}$ across the spike, as if the mode was also decaying there. We see in Fig.\,\ref{fig:q_of_numax} that this effect is small, which can be explained by the fact that the spike is narrow compared to $k^{-1}$. We therefore conclude that it is not necessary to adapt the stellar profiles when calculating the coupling strength since the strong coupling approximation no longer adequately describes the evolution of the coupling when the spike becomes relevant anyway, and the value of the weak coupling is only marginally affected. In fact, the asymptotic coupling values possibly follow a similar evolution as $q_\mathrm{w}$ in response to the chemical discontinuity that causes the spike, with the temporary increase in $q_\mathrm{as}$ as the spike leaves the evanescent zone being most notable for the $1.75\,\mathrm{M}_\odot$-track (cf.\,fourth panel of Fig.\,\ref{fig:q_of_numax_M}). However, this feature is not significant compared to the values' uncertainties and further studies would be needed to investigate this behavior.

\section{Conclusion}\label{s:conclusion}

We empirically verified the validity regimes for the strong and weak coupling approximations on the RGB \citep[as discussed by e.g.\,][]{lit:Pincon2020,lit:Jiang2022a} to coincide with the evanescent zone being fully radiative and convective, respectively. The left two columns of Fig.\,\ref{fig:q_as_vs_q_theo} show that the theoretical couplings agree well with the asymptotic ones in the respective regimes, while $q_\mathrm{w}$ overestimates the coupling when the model lies in the strong coupling regime, and $q_\mathrm{s}$ underestimates the coupling when it is weak. This is also consistent with, e.g., \citet{lit:Jiang2022a}, who find qualitatively similar results (cf.\,Fig.\,5 of that paper) for their order-wise fitting approach. Additionally, we showed that for the transition from one limit to the other, no additional prescription is needed to approximate the fitted values using the stellar structure within the considered mass range. To explore this simplification in coupling strength calculation further, we found the evanescent zone width $W$ (in units of local decay length) as a straightforward, mass-independent parameter of stellar structure to differentiate between the two regimes: For narrow evanescent zones with $W\leq0.4$ e-foldings, the strong coupling (evaluated at frequencies for which the Brunt-Väisälä spike does not lie in the evanescent zone) gives a good approximation for the asymptotic coupling strength; above ($W>0.4$), the weak coupling can be used as an estimate in the mass range considered in this paper ($1.00\,\mathrm{M}_\odot\leq M\leq 2.00\,\mathrm{M}_\odot$). This conclusion is supported by the good agreement of $q_\mathrm{sel}$ (introduced in Eq.\,(\ref{eq:q_sel})) with $q_\mathrm{as}$ as shown in the bottom right panel of Fig.\,\ref{fig:q_as_vs_q_theo}. It adds to the observation by \citet{lit:Jiang2022a}, who also found that the frequency-fitted coupling strength typically aligns with one of the theoretical values, by adding an independent constraint on which approximation to use.

We emphasize that the applicability of the weak coupling approximation at such small evanescent zone widths is purely empirical and relies on the serendipitous cancellation of different assumptions. We do not expect it to necessarily hold for stellar models with masses outside the range we studied here, especially since we observe the agreement to be worst for the models along the lowest- and highest-mass evolutionary tracks used (cf.\,top right panel of Fig.\,\ref{fig:q_as_vs_q_theo}). Still, these findings can aid future work, since they offer a possibility to estimate the coupling strength in a straightforward way, which is a key ingredient for the asymptotic calculation of mixed modes.

We further demonstrated that it is beneficial to calculate the coupling across a frequency range rather than just for $\nu_\mathrm{max}$ in both limits: Noise appearing in the evaluation of the strong coupling can be treated by taking a suitable average of $q_\mathrm{s}$ over the frequency range; and calculating the weak coupling at frequencies close to the upper limit of the frequency range sampled by observations increases agreement with the values found from the mixed mode pattern (which typically assumes $q_\mathrm{as}$ to be a frequency-independent parameter), as can be seen in the bottom right panel of Fig.\,\ref{fig:q_as_vs_q_theo} from the fact that most theoretical values $q_\mathrm{sel}$ lie left of the one-to-one line in the regime of weak coupling. On the other hand, we find that an explicit treatment (or removal) of the spike in the Brunt-Väisälä frequency caused by the chemical discontinuity left behind by the first dredge-up is not necessary when calculating the coupling strength in an RGB-model.

We also tested the influence of stellar mass on the evolution of the coupling: In the strong coupling regime, $q$ tends to be lower for higher masses, due to the steeper gradient of the modified Brunt-Väisälä frequency. Also, the mass affects the evolution of $\nu_\mathrm{max}$ in relation to the evanescent zone width $W$, which is more relevant for the calculation of the coupling strength. Apart from the stronger suppression on the early RGB, we found the evolution of $q$ as a function of $W$ to be almost independent of mass or metallicity.

\begin{acknowledgements}
      We thank Chen Jiang for refereeing this paper and helping increase its clarity and rigor with useful comments. We also thank Nicholas Proietti for contributing ideas to the start of the project. The research leading to the presented results has received funding from the European Research Council Consolidator Grant DipolarSound (grant agreement No. 101000296).\\
      Reproduced with permission from Astronomy \& Astrophysics, © ESO
\end{acknowledgements}

\bibliography{aa55255-25.bib}

\begin{appendix}

\section{Strong coupling approximation}\label{ap:grad_Takata}

In the strong coupling approximation, the evanescent zone is assumed to be so narrow that the behavior of a structure perturbation near the zeros of the asymptotic wave vector $k^2$ (cf.\,Eq.\,(\ref{eq:dispersion_relation})) becomes relevant to the overall behavior of the mode. Since the oscillation equations show singularities at these points, \citet{lit:Takata2016a} introduces a change of coordinates such that the dependent variables become complex and solutions can be derived. The corresponding independent (radius) variable $s$ is defined by
\begin{align}
    r\mapsto s=\ln\left(\frac{r}{\sqrt{r_Sr_N}}\right)\,,
    \label{eq:definition_s}
\end{align}
where $r_{S,N}$ are the turning points where $\omega^2=\hat{S}^2,\hat{N}^2$, respectively. The identification with $r_\mathrm{in,out}$ depends on the choice of oscillation frequency and model, since they determine whether $\omega^2$ crosses $\hat{S}^2$ or $\hat{N}^2$ first. On the RGB, typically $r_\mathrm{in}=r_N,\,r_\mathrm{out}=r_S$ (cf.\,Fig.\,\ref{fig:propagation_diagram}). The radius $r_0$ as it appears in Eq.\,(\ref{eq:T_strong}) is defined by $s(r_0)=0$.

Upon relating the solutions in this frame of coordinates back to physical quantities and using an order-of-magnitude argument that explicitly requires the coupling to be strong, \citet{lit:Takata2016a} finds expression~(\ref{eq:T_strong}) for the transmission coefficient, where
\begin{align}
    \nabla^2(r_0)=\frac{\pi}{2}\left[\sqrt{\frac{s_0^2-s^2}{\mathcal{P}\mathcal{Q}}}\left(\frac{\mathrm{d}\ln\mathfrak{c}}{\mathrm{d}s}\right)^2\right]_{s=0}\,.
    \label{eq:grad_Takata}
\end{align}
For a wide evanescent zone, the shape of the wave function would change and this term would no longer appear in leading order and hence, the suppression of $T_\mathrm{s}$ is overestimated.

The reference coordinate $s_0$ is given by
\begin{align}
	s_0 \coloneqq s(r_S) = \ln\left(\sqrt{\frac{r_S}{r_N}}\,\right) = -s(r_N)\,,
\label{eq:definition_s0}
\end{align}
and can be both positive or negative. The gradient $\frac{\mathrm{d}\ln\mathfrak{c}}{\mathrm{d}s}$ contains multiple terms:
\begin{align}
    \frac{\mathrm{d}\ln\mathfrak{c}}{\mathrm{d}s}=\frac{1}{2}\left(\frac{s_0}{s_0^2-s^2}-(\mathcal{V}-\mathcal{A}-J)+\frac{1}{2}\left(\frac{\mathrm{d}\ln\mathcal{P}}{\mathrm{d}\ln r}-\frac{\mathrm{d}\ln\mathcal{Q}}{\mathrm{d}\ln r}\right)\right)\,.
    \label{eq:dlnc_ds}
\end{align}
In this expression, 
\begin{align}
    \mathcal{V} &= -\frac{1}{\Gamma_1J}\frac{\mathrm{d} \ln p}{\mathrm{d} \ln r}\quad\text{and}\label{eq:V_Takata}\\
	\mathcal{A} &= \frac{1}{J}\left(\frac{1}{\Gamma_1}\frac{\mathrm{d} \ln p}{\mathrm{d} \ln r}-\frac{\mathrm{d} \ln\rho}{\mathrm{d} \ln r}\right)\label{eq:A_Takata}
\end{align}
relate to the modified Lamb and Brunt-Väisälä frequency in the new coordinate frame, respectively, and therefore
\begin{align}
    \mathcal{P} &= 2J-\lambda\mathcal{V}\label{eq:P_Takata}\quad\text{and}\\
	\mathcal{Q} &= J-\frac{\mathcal{A}}{\lambda}\label{eq:Q_Takata}
\end{align}
correspond to the two bracketed terms in the asymptotic dispersion relation Eq.\,(\ref{eq:dispersion_relation}) in terms of a dimensionless frequency
\begin{align}
    \lambda     &= \frac{\omega^2r}{g}\label{eq:lambda_Takata}\,.
\end{align}
This means that $\mathcal{P}$ and $\mathcal{Q}$ are zero at $r_{S,N}$, respectively, and therefore their logarithmic derivatives appearing in Eq.\,(\ref{eq:dlnc_ds}) diverge at the limits of the evanescent zone, which demonstrates that $\frac{\mathrm{d}\ln\mathfrak{c}}{\mathrm{d}s}$ needs to span a vast range of values within this narrow region. Therefore, the noise created by numerically evaluating the many gradients of stellar structure would affect the value of the transmission coefficient (and hence $q_\mathrm{s}$) strongly.

To mitigate this, we used two equilibrium (static) equations of stellar structure to evaluate as many derivatives analytically as possible: The continuity equation
\begin{align}
    \frac{\mathrm{d} m}{\mathrm{d} r} = 4\pi r^2\rho\,,
    \label{eq:continuity}
\end{align}
where $m(r)$ is the mass contained within radius $r$; and the equation of hydrostatic equilibrium
\begin{align}
    \frac{\mathrm{d} p}{\mathrm{d} r}&= -\frac{Gm}{4\pi r^4}\frac{\mathrm{d} m}{\mathrm{d} r} = -\frac{Gm\rho}{r^2}\,.
    \label{eq:hydrostatic_eq}
\end{align}
Using these, we can write:
\begin{align}
	\mathcal{V} &= \frac{1}{\Gamma_1J}\frac{Gm\rho}{pr}\\
	\mathcal{A} &= \frac{1}{J}\frac{Gm\rho}{pr}\underbrace{\left(\frac{\mathrm{d} \ln\rho}{\mathrm{d} \ln p}-\frac{1}{\Gamma_1}\right)}_{=: \Gamma_\mathcal{A}^{-1}} \\
	\frac{\mathrm{d} J}{\mathrm{d} r} &= \frac{\rho}{r\rho_{in}}\left(\frac{Gm}{r}\frac{\rho}{p}\frac{\mathrm{d} \ln\rho}{\mathrm{d} \ln p}-3J\right)\\
    \frac{\mathrm{d}\ln\mathcal{P}}{\mathrm{d}\ln r} &= \frac{r}{\mathcal{P}}\left[2\frac{\mathrm{d} J}{\mathrm{d} r}-\omega^2\frac{\rho r^2}{p}\frac{1}{\Gamma_1J}\left(\frac{Gm\rho}{pr^2}\left(1-\frac{\mathrm{d}\ln\rho}{\mathrm{d}\ln p}\right)\right.\right.\nonumber\\&\qquad\left.\left.+\frac{2}{r}+\Gamma_1\frac{\mathrm{d}}{\mathrm{d} r}\left[\frac{1}{\Gamma_1}\right]-\frac{1}{J}\frac{\mathrm{d} J}{\mathrm{d} r}\right)\right]\\
	\frac{\mathrm{d}\ln\mathcal{Q}}{\mathrm{d}\ln r} &= \frac{r}{\mathcal{Q}}\left[\frac{\mathrm{d} J}{\mathrm{d} r}-\frac{G^2}{\omega^2}\frac{\rho m^2}{pr^4}\frac{1}{\Gamma_\mathcal{A}J}\left(\frac{Gm\rho}{pr^2}\left(1-\frac{\mathrm{d}\ln\rho}{\mathrm{d}\ln p}\right)+\frac{8\pi\rho r^2}{m}\right.\right.\nonumber\\&\qquad\left.\left.-\frac{4}{r}+\Gamma_\mathcal{A}\frac{\mathrm{d}}{\mathrm{d} r}\left[\frac{1}{\Gamma_\mathcal{A}}\right]-\frac{1}{J}\frac{\mathrm{d} J}{\mathrm{d} r}\right)\right]
\end{align}
The remaining two gradients of the adiabatic index $\Gamma_1$ and polytropic index $\frac{\mathrm{d}\ln p}{\mathrm{d}\ln\rho}$ still need to be computed numerically as differential quotients. Since both quantities only vary slowly across the stellar profile, this is feasible -- however, these terms still are the main source of noise, especially since the polytropic index being calculated from tabulated values for the equation of state is already subject to numerical noise itself.

\end{appendix}

\end{document}